\documentclass[12pt,a4paper]{article}

\usepackage{latexsym}
\usepackage{wasysym}

\title{Sensitivity of high precision Michelson-Morley experiments to tilting of their setups}

\author{Ll.\ Bel \footnote{wtpbedil@lg.ehu.es}\\
\emph{Fisika Teorikoa, Euskal Herriko Unibertsitatea}, \\
\emph{P.K. 644, 48080 Bilbo, Spain}
}

\begin{document}
\maketitle

\begin{abstract}

We describe the effects to be expected of unwanted or voluntary deviations from the vertical of the axis of the active rotation of modern high precision experiments of the Michelson-Morley type. The theoretical description that we use is a particular implementation of the Principle of free mobility.

\end{abstract}

\section*{Introduction}

The first high precision experiment of the Michelson-Morley type was done by Brillet and Hall at Boulder in 1979 \cite{Brillet-Hall} and three new ones have been completed recently at Dusseldorf by Antonini et al. \cite{Antonini}, at Perth by Stanwix et al. \cite{Stanwix} and at Berlin by Herrmann et al. \cite{Herrmann}. A simplified description of these experiments is the following: the round-trip speed of light propagating along a horizontal path, in vacuum or a sapphire crystal, is continuously monitored while this direction of propagation is actively rotated with, say, angular velocity $\omega$. Actually what is monitored is the frequency of resonance of a cavity which is compared to a fixed frequency of reference, \cite{Brillet-Hall} and \cite{Herrmann}, or to the frequency of a second rotating cavity, \cite{Antonini} and \cite{Stanwix},  at right angles with respect to the first one.

The purpose of a Michelson-Morley experiment is to check the isotropy of the round trip speed of light on the surface of the Earth, or, if there is some anisotropy, to describe it trying to identify the cause. If the speed of light were isotropic, a perfect Michelson-Morley experiment, but none is, would show the isotropy whatever the orientation of the setup. On the other hand if there is some anisotropy, whatever the cause, the output of the experiment will be sensitive to the orientation of the setup. For this reason no experiment of the Michelson-Morley type could be considered completely satisfactory without a careful analysis of the results of voluntary tilts of the axis of rotation, both from the experimental point of view and from the theoretical one, since this might be the unique way to discriminate between a spurious signal from an interesting real one. Unfortunately, to our knowledge, this has not been done up to now.

	To do such an analysis from the theoretical point of view requires to consider particular models to describe potential anisotropies. In this paper we consider a model which is a particular implementation of the Principle of free mobility, a principle that demands the possibility of describing ideal rigid bodies.
	
	The first section is a remainder of the simplified standard model of the Earth and its exterior gravitational field in the framework of linearized  General relativity theory.
	
	The second second section is a remainder of the particular implementation of the Principle of free mobility that we are considering and lays down the main formulas to be used in our calculation.
	
	The third section is a remainder of the signal $c(\vec n)$ to be analyzed, i.e. the function describing the values of the speed of light at some location depending on the direction of propagation, and introduces the three orthonormal basis to be used to discuss the tilt effects. 

	The final section contains the formulas describing the principal parts of the signal, allowing for the possibility that the axis of rotation of the Michelson-Morley experiment is tilted an angle $\delta$ away from the vertical of the location. 
		
	The general form of the signal is:
	
\begin{equation}
c(\vec n):=1+a_0+a_2\cos 2\omega\tau+b_2\sin 2\omega\tau	\nonumber
\end{equation}
$\tau$ being local time and $a_2$ and $b_2$ the two quantities that are accessible to the experimentalists. And it follows from our formulas that:

i) When the tilt angle $\delta$ is zero then, choosing an appropriate origin to measure the angle $\omega\tau$, one has $b_2=0$ and $a_2$ of the order of $10^{-13}$ at intermediate latitudes. This was already known \cite{Bel-Molina}.

ii) For values of $\delta\leq 10^{-3}$ the contributions of the tilt to modifications of $b_2$ and $a_2$ are generically of order $10^{-15}$
or smaller, and more importantly:

iii) There always exist a tilt around the value $\delta=0.06$ rad, i.e. $\delta=3.4^\circ$, and an appropriate orientation such that $b_2=0$ and $a_2=0$.

\section{A simplified model of the gravitational field of the Earth}

We shall use the simplified  model of the  Earth that takes into account its mass $M$, its quadrupole moment per unit mass $J_2$, its proper angular velocity $\Omega$ and that approximates its shape as that of an equipotential ellipsoid of revolution with equatorial radius $a$ and polar radius $b$.

We shall use units such that:

\begin{equation}
\label{1.1}
c=1, \quad G=1
\end{equation} 
The afore-mentioned quantities will have then the following approximate values, m meaning meter:

\begin{equation}
\label{1.2}
M=4.4\times 10^{-3}\,\hbox{m}, \quad J_2=10^{-3}
\end{equation}
\begin{equation}
\label{1.3}
a=63.78\times 10^4\,\hbox{m}, \quad b=63.56\times 10^4\,\hbox{m}
\end{equation}
and:
\begin{equation}
\label{1.4}
\Omega=2.43\times 10^{-13}\,\hbox{m${}^{-1}$}
\end{equation}

We consider the following dimensionless quantities:

\begin{equation}
\label{1.5}
\frac{M}{a}=7\times 10^{-10}, \quad \Omega^2a^2=2\times 10^{-11}, \quad
\frac{M}{a}J_2=7\times 10^{-13}
\end{equation}
Products and squares of quantities of the order of the above ones will be neglected in the calculations to follow.

With these assumptions the linear approximation of the line-element of the exterior gravitational field of the Earth can be written, with an obvious preliminary meaning of the polar coordinates $t$, $\hat r$, $\hat\theta$ and $\varphi$ as:

\begin{equation}
\label{1.7}
ds^2=-\Psi^2+d\hat s^2
\end{equation}
where:

\begin{equation}
\label{1.7.1}
\Psi=-(1-U_G-U_\Omega)dt+\Omega \hat r^2\sin^2\hat\theta d\varphi
\end{equation}
with:

\begin{equation}
\label{1.8}
U_G=\frac{M}{\hat r}\left(1+\frac{J_2a^2}{\hat r^3}(1-3\cos^2\hat\theta)\right), \quad 
U_\Omega=\frac12\Omega^2\hat r^2\sin^2\hat\theta
\end{equation}
and:

\begin{equation}
\label{1.9}
d\hat s^2=(1+2U_G)(d\hat r^2+\hat r^2d\hat\theta^2)
+(1+2U_G+2U_\Omega)\hat r^2\sin^2\hat\theta
\end{equation}

%% Mention also the order of magnitude of epsilon=1-(a/b)^2

\section{The Principle of free mobility}

While the interpretation of $\Psi$ in (\ref{1.7.1}) poses no problem and establishes the connection between Einstein's theory of gravitation and Newtonian physics concepts, on the contrary the interpretation of the quadratic form (\ref{1.9}) should in our opinion be put to debate. In fact, since the Riemannian curvature of this metric is not constant the usual interpretation that identifies it as describing the geometry of space makes impossible to define rigid structures without which the whole science of metrology breaks down. 

As an element to this debate we have offered before the idea that Relativity theories should not renounce to the Principle of free mobility of ideal rigid bodies and this requires the geometry of space to have constant curvature, or in particular to be Euclidean which is the only case that we consider here. More precisely our proposition is to accept that the geometry of space is described by a proper Riemannian metric:

\begin{equation}
\label{2.1}
d\bar s^2=\bar g_{ij}dx^idx^j
\end{equation}
satisfying the following conditions:

i)the metric is Euclidean, i.e. its Riemann tensor is zero: 

\begin{equation}
\label{2.2}
\bar R_{ijkl}=0
\end{equation} 
and:

ii) satisfies the supplementary condition:

\begin{equation}
\label{2.3}
(\hat\Gamma^i_{jk}-\bar\Gamma^i_{jk})\hat g^{jk}=0
\end{equation}
where $\bar\Gamma^i_{jk}$ and $\hat\Gamma^i_{jk}$ are respectively the Christoffel symbols of $d\bar s^2$ and $d\hat s^2$ written:

\begin{equation}
\label{2.3.1}
d\hat s^2=\hat g_{ij}dx^idx^j
\end{equation}
and where $\hat g^{ij}$ is the inverse of $\hat g_{ij}$. Notice that both conditions, being 3-dimensional tensor equations, are intrinsic to the Killing congruence defining the frame of reference co-moving with the Earth. And that while i) is a general condition to implement the Principle of free mobility and is independent of what metric we start with, condition ii) ties any particular expression of it to a particular expression of the Euclidean metric. The choice of this second condition is of course open to discussion. The choice that we made here is at the moment our best educated guess.
 
From (\ref{2.3}) it follows that if, once the metric (\ref{2.1}) has been obtained, Cartesian coordinates $x^i$ are used then these coordinates are quo-harmonic coordinates\footnote{We say quo-harmonic instead of harmonic coordinates to underline the fact that they are not harmonic coordinates of the space-time metric} of $d\hat s^2$. This provides a method to obtain the metric (\ref{2.1}) directly by performing the approximate coordinate transformation to write $d\hat s^2$ in quo-harmonic coordinates, a process that involves solving a system of Poisson equations. This method was outlined in \cite{Bel-Molina} and will not be repeated here.

After this coordinate transformation has been performed, using true polar coordinates, the metric (\ref{2.1}) becomes: 

\begin{equation}
\label{2.4}
d\bar s^2=\delta_{ij}\theta^i\theta^j 
\end{equation}
where :

\begin{equation}
\label{2.5}
\theta^1=dr, \quad \theta^2=rd\theta, \quad \theta^3=r\sin\theta d\varphi
\end{equation}
and the metric $d\hat s^2$ is: 
 
\begin{equation}
\label{2.6}
d\hat s^2=(\delta_{ij}+h_{ij})\theta^i\theta^j
\end{equation}
with: 

\begin{eqnarray}
\label{2.6.1}
h_{11}=\frac{2M}{r}\left(1+\frac15\frac{a^2}{r^2}\right)
-\frac{2MJ_2a^2}{r^3}\left(1+\frac67\frac{a^2}{r^2}-\left(1+\frac97\frac{a^2}{r^2}\right)\sin^2\theta\right) \nonumber \\
+\frac32\Omega^2 r^2\sin^4\theta 
+\frac25\frac{Ma^2}{r^3}\sigma_1+12\frac{Ma^4J_2}{r^5}(1-3\cos^2\theta)\sigma_2
\end{eqnarray}

\begin{eqnarray}
\label{2.7}
h_{22}=\frac{3M}{r}\left(1-\frac{1}{5}\frac{a^2}{r^2}\right)
-\frac{MJ_2a^2}{r^3}\left(3-\frac67\frac{a^2}{r^2}-\left(\frac{11}{2}-\frac{3}{2}\frac{a^2}{r^2}\right)\sin^2\theta\right) \nonumber \\
+\frac32\Omega^2 r^2\sin^2\theta\cos^2\theta 
-\frac15\frac{Ma^2}{r^3}\sigma_1-3\frac{Ma^4J_2}{r^5}(3-7\cos^2\theta)\sigma_2
\end{eqnarray}

\begin{eqnarray}
\label{2.8}
h_{33}=\frac{3M}{r}\left(1-\frac{}{15}\frac{a^2}{r^2}\right)
-\frac{MJ_2a^2}{r^3}\left(3-\frac67\frac{a^2}{r^2}-\left(\frac{9}{2}-\frac{15}{14}\frac{a^2}{r^2}\right)\sin^2\theta\right)\nonumber \\
+\frac32\Omega^2 r^2\sin^2\theta 
-\frac15\frac{Ma^2}{r^3}\sigma_1-3\frac{Ma^4J_2}{r^5}(1-5\cos^2\theta)\sigma_2
\end{eqnarray}

\begin{equation}
\label{2.9}
h_{12}=\left(\frac{2MJ_2a^2}{r^3}\left(1-\frac67\frac{a^2}{r^2}\right)+\frac32\Omega^2 r^2\sin^2\theta-24\frac{Ma^4J_2}{r^5}\sigma_2\right)\sin\theta\cos\theta
\end{equation}
$\sigma_1$ and $\sigma_2$ being two constants of integration that should be fixed by matching the exterior solution above to the interior solution corresponding to a particular model for the interior of the Earth, as it was done in Sect~3 of \cite{Bel-Molina},where it was assumed that its density was constant\footnote{Notice that in this reference $a$ is the polar radius and the equatorial radius is named $R$}. In this particular case these constants are:

\begin{equation}
\label{2.10}
\sigma_1=0, \quad \sigma_2=0
\end{equation}

\section{Modern high precision Michelson-Morley experiments}

Let us consider a light ray that leaves from a point $P$ with coordinates $x^k$ at proper time $\tau$ and reaches a neighboring point $P+dP$ with coordinates $x^k+dx^k$ where it is reflected back and reaches the point $P$ again at proper time $\tau+2d\tau$. The world-line of a light ray propagating in vacuum\footnote{Ref. \cite{Crucial} describes how to deal with the general case.} being a null geodesic of the space-time metric we always have:

\begin{equation}
\label{3.1}
\frac{1}{d\tau}\sqrt{\hat g_{ij}dx^idx^j}=1
\end{equation}
A result that justifies the much repeated assertion that the speed of light is always $1$ without reminding that this assertion is based on an interpretation of the quadratic form $d\hat s^2$ as describing the geometry of space, an interpretation incompatible with the Principle of free mobility.

If we accept that the physical distance from $P$ to $P+dP$ instead of being $d\hat s$ is:

\begin{equation}
\label{3.2}
d\bar s=\sqrt{\bar g_{ij}dx^idx^j}
\end{equation}
then the speed of light for a short round-trip travel becomes:

\begin{equation}
\label{3.3}
c(\vec n)=\frac{d\bar s}{d\tau}=\sqrt{\frac{\bar g_{ij}dx^idx^j}{\hat g_{kl}dx^kdx^l}}
\end{equation} 
where $\vec n$ is the unit vector in the direction of propagation of light.
Or using (\ref{2.4}) and (\ref{2.6}):

\begin{equation}
\label{3.4}
c(\vec n)=1-\frac12(h_{11}(n^1)^2+h_{22}(n^2)^2+h_{33}(n^3)^2+2h_{12}n^1n^2)
\end{equation}
$n^i$ being the components of $\vec n$ with respect to the orthonormal Euclidean natural basis, the dual of (\ref{2.5}), so that:

\begin{equation}
\label{3.5}
\vec n=n^1\vec e_1+n^2\vec e_2+n^3\vec e_3.
\end{equation}
$\vec e_1$ is in the direction of the polar vector from the center of the Earth to the location of the experiment, $\vec e_2$ is in the direction of the meridian and $\vec e_3$ is in the direction of the geographic parallel. 

To discuss the effect of an unwanted or voluntary tilt of the axis of the active rotation of the direction of propagation of light we need to introduce two other basis besides the dual of (\ref{2.5}). The first one has its first vector $\vec v_1$ in the direction of the vertical. Therefore:

\begin{equation}
\label{3.6}
\vec v_1=\cos\alpha\,\vec e_1+\sin\alpha\,\vec e_2
\end{equation}
where $\alpha$ has to be derived from the equation of the equipotential surface of the Earth: 

\begin{equation}
\label{3.7}
U=\hbox{Const.} \quad \hbox{with} \quad U=U_G+U_\Omega
\end{equation}
We obtain thus:

\begin{equation}
\label{3.8}
\alpha=-\left(\frac{3J_2a^2}{r^2}+\frac{\Omega^2r^3}{M}\right)\sin\theta\cos\theta
\end{equation}
Equivalently, approximating the equation of the surface of the Earth: 

\begin{equation}
\label{3.4.1}
r=\sqrt{b^2\cos^2\theta+a^2\sin^2\theta}
\end{equation}
by:

\begin{equation}
\label{3.4.2}
r\approx a(1-\frac12\epsilon\cos^2\theta)
\end{equation}
with: 

\begin{equation}
\label{3.4.3}
\epsilon=1-\frac{b^2}{a^2} \approx 6.7\times 10^{-3}
\end{equation}
we can write :

\begin{equation}
\label{3.8.1}
\alpha=-\epsilon\sin\theta\cos\theta
\end{equation}

To complete the triad we use:

\begin{equation}
\label{3.9}
\,\vec v_2=-\sin\alpha\,\vec e_1+\cos\alpha\,\vec e_2, \quad \vec v_3=\,\vec e_3
\end{equation}

We define the second triad as follows:

\begin{equation}
\label{3.10}
\,\vec t_1=\cos\delta\,\vec v_1+\sin\delta\,\vec v_2
\end{equation} 
where $\delta$ is the angle of the tilt, and:

\begin{equation}
\label{3.10.1}
\,\vec t_2=\sin\chi\,\vec v_1+\cos\chi\,\vec v_2
\end{equation} 
\begin{eqnarray}
\label{3.11}
\,\vec t_3=-\sin\beta\cos\chi\sin\delta\,\vec v_1
+\sin\beta\sin\chi\sin\delta\,\vec v_2 \nonumber \\
+(\cos\chi\cos\delta-\sin\chi\cos\beta\sin\delta)\,\vec v_3
\end{eqnarray}
where $\beta$ is the angle of the projection of $\,\vec t_1$ on the plane $\,\vec v_2-\,\vec v_3$ and where:

\begin{equation}
\label{3.12}
\chi=-\arctan(\cos\beta\tan\delta)
\end{equation}  
has been chosen so that $\,\vec t_2$ lies on the plane $\,\vec v_1-\,\vec v_2$

With the preceding definitions the direction of propagation of light will be assumed to be given by:

\begin{equation}
\label{3.13}
\,\vec n=\cos\omega\tau\,\vec t_2+\sin\omega\tau\,\vec t_3
\end{equation}  

Beyond this point the theoretical prediction about the outcome of the experiment, and the effects of any particular tilt, have to be derived from particular models.
The next section will conclude this paper by considering the model presented in Sec.~2 based primarily on the Principle of free mobility.

\section{Concluding results}

Besides the approximations already mentioned the results below derived by a straight-forward calculation of $c(\,\vec n)$ in (\ref{3.4}) assume that $\delta$ is a small quantity and that terms of order $\delta^3$  or smaller can be neglected. 

We write the final results assuming (\ref{2.10}) and including only the leading terms for each of the coefficients in the expressions: 

\begin{equation}
\label{4.4}
c(\vec n):=1+a_0+a_2\cos 2\omega\tau+b_2\sin 2\omega\tau
\end{equation}
where:

\begin{eqnarray}
\label{4.5}
a_0&=&a_{00}+a_{01}\delta+a_{02}\delta^2  \\
a_2&=&a_{20}+a_{21}\delta+a_{22}\delta^2  \\ 
b_2&=&a_{20}+b_{21}\delta+b_{22}\delta^2
\end{eqnarray}
They are the following:

\begin{eqnarray}
\label{4.6}
a_{00}&=&-\frac75\frac{M}{a}  \\
\label{4.6.1}
a_{01}&=&-\frac15\frac{\epsilon M}{a}\sin\theta\cos\theta  \\
\label{4.6.2}
a_{02}&=&-\frac35\frac{M}{a}  
\end{eqnarray}
and:
\begin{eqnarray}
\label{4.7}
a_{20}&=&-\frac17\frac{J_2M}{a}\sin^2\theta+\frac38\Omega^2 a^2\sin^4\theta  \\
\label{4.7.1}
b_{20}&=& 0  \\
\label{4.7.2}
a_{21}&=&\left(-\frac15\frac{\epsilon M}{a}
+\frac34\Omega^2a^2\sin^2\theta+\frac17\frac{J_2M}{a}\right)\sin\theta\cos\theta\cos\beta  \\
\label{4.7.3}
b_{21}&=& \left(-\frac15\epsilon\frac{M}{a}+\frac34\Omega^2a^2\sin^2\theta+\frac17\frac{J_2M}{a}\right)\sin\theta\cos\theta\sin\beta  \\
\label{4.7.4}
a_{22}&=&-\frac15\left(\frac12-\cos^2\beta\right)\frac{M}{a}  \\
\label{4.7.5}
b_{22}&=& \frac15\frac{M}{a}\sin\beta\cos\beta  
\end{eqnarray} 
The terms that have been excluded from (\ref{4.6})-(\ref{4.7.5}) would lead to terms of order $10^{-15}$ or smaller with tilts of the order of $\delta= 10^{-2}$. 

The term $a_0$ is a time independent term that affects the frequency of resonance of the rotating cavity as well as the frequency of reference and it is therefore irrelevant to the Michelson-Morley experiments that we are considering. 

The signal (\ref{4.4}) does not contain any other time dependent Fourier amplitude besides the one at $2\omega\tau$. This is part of the prediction, and therefore choosing an orientation that minimizes the first Fourier amplitude at $\omega\tau$ is in principle a good strategy to check the alignment of the axis of the active rotation with the vertical of the location\footnote{A.~Brillet, private communication}.  

If the axis of the active rotation is aligned with the vertical of the location of the experiment then the prediction, as already derived in \cite{Bel-Molina}, is given  by $a_{20}$ and $b_{20}$ in (\ref{4.7}) and (\ref{4.7.1}), so that:

\begin{equation}
\label{4.7.6}
a_2=\frac38\Omega^2 a^2\sin^4\theta -\frac17\frac{J_2M}{a}\sin^2\theta, \quad  b_2=0
\end{equation}

From (\ref{4.7})-(\ref{4.7.5}) it follows that the system of equations:

\begin{equation}
\label{4.8}
a_2(\delta_0, \cos\beta_0)=0, \quad   b_2(\delta_0, \cos\beta_0)=0
\end{equation}
can be solved consistently yielding:

\begin{eqnarray}
\label{4.9}
\delta_0&=&\sqrt{\frac{15}{4}\frac{\Omega^2a^3}{M}\sin^2\theta-\frac{10}{7}J_2}\ \sin\theta \\
\label{4.9.1}
\cos\beta_0&=&-\frac{1}{\delta_0}\left(\frac{15}{4}\frac{\Omega^2a^3\sin\theta^2}{M}+\frac57 J_2-\epsilon\right)\sin\theta\cos\theta
\end{eqnarray}
The existence of this solution for $\delta_0$ and $\beta_0$ may lead to a strategy to prove the existence of the effect (\ref{4.7.6}) by nullifying it! This fact depends crucially on taking into account the monopole contribution of the gravitational field of the Earth. 

To underline the remarks above we mention below some of the relevant numeric results corresponding to the co-latitude of Boulder:
 
\begin{equation}
\label{4.10}
\theta=(50^\circ/180^\circ)\pi\approx 0.87\hbox{\,rad}
\end{equation}
We get then:
\begin{equation}
\label{4.11}
\sqrt{a_2^2+b_2^2}=2.5\times 10^{-13}, \quad \arctan\frac{b_2}{a_2}=0^\circ
\end{equation}
The effect is mainly due to the Earth rotation, and this explains why models that ignore the gravitational field yield a similar result \cite{Bel}, \cite{Klauber}.

Brillet and Hall did mention an effect of this order with:

\begin{equation}
\label{4.11.1}
\sqrt{a_2^2+b_2^2}=2\times 10^{-13} , \quad \arctan\frac{b_2}{a_2}=-25^\circ
\end{equation} 
without identifying the geographical direction of reference. They suggested that it could be spurious and their analysis was then restricted to discuss the residuals from this values that had generically an amplitude of the order of $10^{-15}$. This is the order of magnitude of the terms that have been neglected from (\ref{4.7}) and from tilts of the order of $\delta=10^{-3}$. This is also the order of magnitude of the effect that is mentioned by the authors of Refs. \cite{Antonini}, \cite{Stanwix} and \cite{Herrmann}, who do not mention any systematic effect comparable to (\ref{4.11.1}). We suggest that if this effect was not seen by them might be due to the fact that their data analysis bypassed the first step of the data analysis done by Brillet and Hall, very clearly summarized in the Fig~2 of their paper.  

The tilt and the orientation of the setup at Boulder that would have nullified the outcome of the Michelson-Morley experiment was according to our model and analysis:

\begin{equation}
\label{4.12}
\delta_0=0.06 \hbox{\,rad}, \quad \beta_0=\arccos(-0.014) 
\end{equation}
This important tilt, $\delta=3.4^\circ$, is again a reflect of the dominant contribution of the rotation of the Earth in Eq. (\ref{4.7}).

\end{document}